\documentclass[pra,aps,reprint]{revtex4-1}
\usepackage{amsthm}
\usepackage{amsmath}
\usepackage{amssymb}
\usepackage{xcolor}
\usepackage[linkbordercolor=blue,
citebordercolor=blue]{hyperref}
\usepackage{graphicx}
\usepackage{braket}
\newtheorem{theorem}{Theorem}
\newcommand{\vc}[1]{{\mathbf{#1}}}

\begin{document}
\title{Unitary quantum perceptron as efficient universal approximator}

\author{E. Torrontegui}
\email{eriktorrontegui@gmail.com}
\address{Instituto de F\'{\i}sica Fundamental IFF-CSIC, Calle Serrano 113b, 28006 Madrid, Spain}
\author{J. J. Garc{\'\i}a-Ripoll}
\address{Instituto de F\'{\i}sica Fundamental IFF-CSIC, Calle Serrano 113b, 28006 Madrid, Spain}

\begin{abstract}  	
We demonstrate that it is possible to implement a quantum perceptron with a sigmoid activation function as an efficient, reversible many-body unitary operation. When inserted in a neural network, the perceptron's response is parameterized by the potential exerted by other neurons. We prove that such a quantum neural network is a universal approximator of continuous functions, with at least the same power as classical neural networks. While engineering general perceptrons is a challenging control problem --also defined in this work--, the ubiquitous sigmoid-response neuron can be implemented as a quasi-adiabatic passage with an Ising model. In this construct, the scaling of resources is favorable with respect to the total network size and is dominated by the number of layers. We expect that our sigmoid perceptron will have applications also in quantum sensing or variational estimation of many-body Hamiltonians.
\end{abstract}  	

\maketitle

Quantum computing and machine learning are two computing paradigms that fight the limitations of procedural programming. While the first one is based on a \textit{physically} different model of computation, the second one reuses von Neumann architectures to build sophisticated approximation models that outperform traditional algorithms. Quantum machine learning merges ideas from both paradigms\ \cite{Biamonte2017,Schuld2015a}, to create new quantum algorithms such as engine ranking\ \cite{Garnerone2012}, data fitting\ \cite{Wiebe2012}, autoencoders\ \cite{Lamata2017,Romero2017}, or autonomous agents\ \cite{Dunjko2016}.

\begin{figure}[h]
  \includegraphics[width=0.75\linewidth]{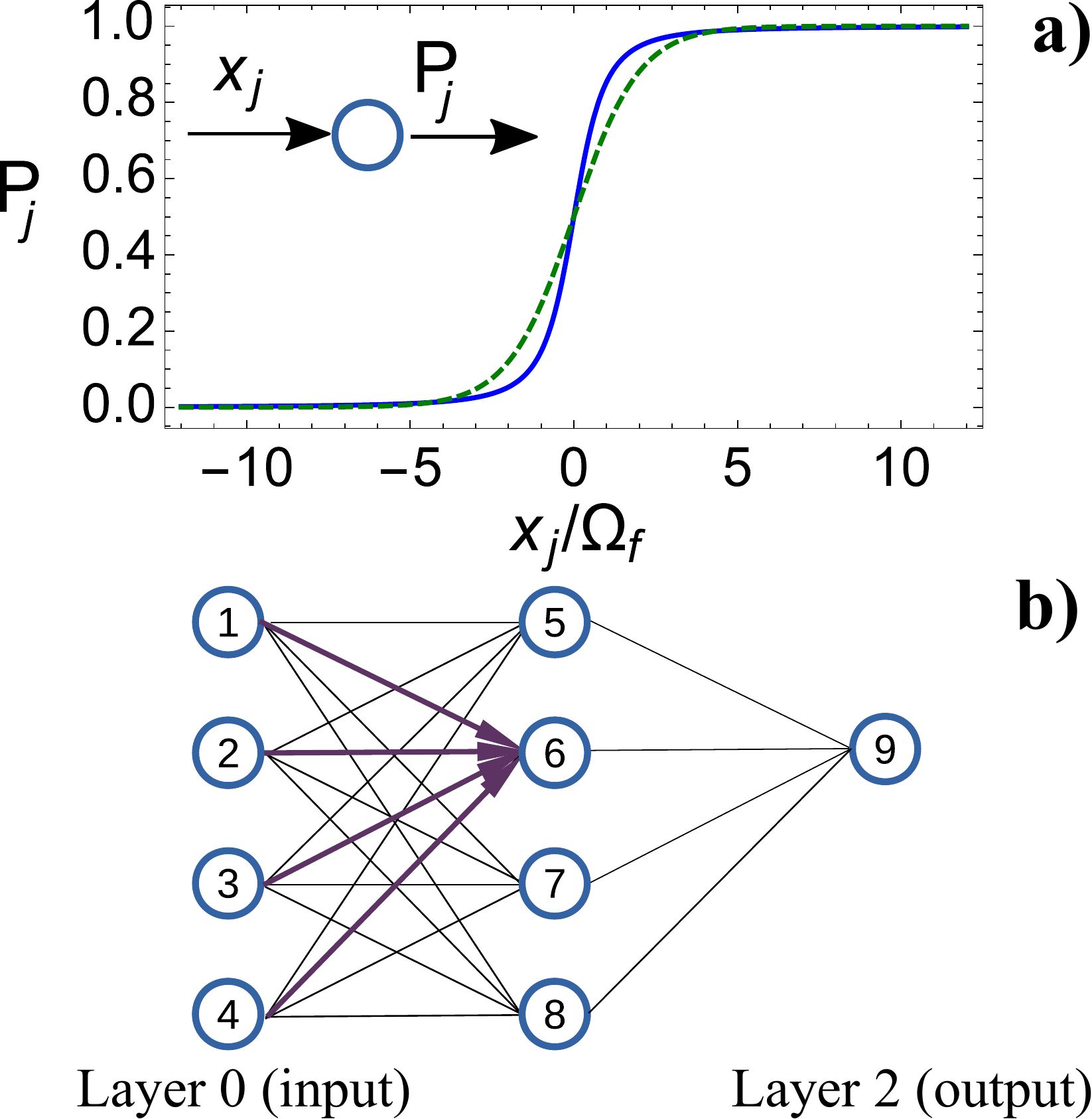}
  \caption[fig:network]{(a) Quantum perceptron as a qubit that excites coherently according to \eqref{eq:quantum-neuron} with a probability $P_j=\frac{1}{2}(1+\braket{\hat\sigma^z_j})=f(x_j)$ that grows nonlinearly with the activation potential $x_j$. (b) When this perceptron is integrated in a feed-forward neural network, the potential depends on neurons in earlier layers, e.g. $x_{6}=\sum_{k=1}^{4} w_{6,k}\hat\sigma^z_k+\theta_{6}$.}
  \label{fig:network}
\end{figure}

In this work we challenge the notion of quantum neural networks, a term claimed by quantum machine learning works\ \cite{Toth1996, Narayanan2000, Altaisky2001, Gupta2001, Andrecut2002, Fei2002, Zhou2012, DaSilva2012, Liu2013, Altaisky2014, DaSilva2016, Wan2017}, which is far from settled\ \cite{Schuld2014}. A feed-forward neural network is made of perceptrons\ \cite{Hopfield1984} that generate signals, $s_j=f(x_j)$, as a nonlinear response to the weighted influence of other neurons, with some intrinsic biases $x_j=\sum_{k<j} w_{jk}s_{k}-\theta_j$ [cf. Fig.\ \ref{fig:network}b]. Classical feed-forward networks are universal approximators of continuous functions\ \cite{Hornik1989} and are trained using reduced information to solve complex problems. A quantum analog of neural network faces the need of (i) encoding the network in a Hilbert space, (ii) defining a physical operation for the neuron activation, (iii) designing an algorithm to train the network and, most important, (iv) finding real-world applications of the quantum version.

We address these problems with a quantum perceptron that is a qubit with a nonlinear excitation response to an input field [cf. Fig.\ \ref{fig:network}a]
\begin{equation}
  \label{eq:quantum-neuron}
  \hat U_j(\hat{x}_j;f)\ket{0_j} = \sqrt{1 - f(\hat{x}_j)}\ket{0_j} + \sqrt{f(\hat{x}_j)} \ket{1_j},
\end{equation}
In a feed-forward network, the perceptron gate is conditioned on the field generated by neurons in earlier layers, $\hat{x}_j = \sum_{k<j} w_{jk}\hat\sigma^z_k - \theta_j$, with similar weights $w_{jk}$ and biases $\theta_j$ as classical networks. This allows us to prove that a network based on this perceptron is a universal approximator of arbitrary continuous functions. We also prove that the perceptron gate $\hat U_j(\hat{x}_j;f)$ has an efficient hardware implementation as a quasiadiabatic passage in an Ising model of interacting spins, with an implementation time that scales favorably $\mathcal{O}(L\times \log(\varepsilon/N)/\Omega_f)$, with the number of layers $L$, number of neurons $N$, gate error $\varepsilon$ and activation step size $\Omega_f$ [cf. Fig.\ \ref{fig:network}a]. In addition to reproducing classical neural networks using quantum states, other applications of this perceptron include the design of multiqubit conditioned quantum gates, or the design of more general perceptrons with sophisticated response functions that can be applied in quantum sensing or classification of quantum states. Our perceptron is intimately related to a recent proposal by Cao \textit{et al.}\ \cite{Cao2017}, which implements the nonlinear activation of a qubit using repeat-until-success quantum gates. As discussed later, our perceptron shares the same potential applications with various advantages: universality, scaling of resources, avoidance of phase wrapping (works for arbitrarily large $|x|$) and utility for general nonlinear sensing.

\section{Classical neural networks}%
Classical neurons are modeled as a mathematical system which may become \emph{active} $(s=1)$ or remain \emph{resting} $(s=0)$, as a response to the state of other $n$ neurons. The neuron activation or \emph{perceptron}\ \cite{Rosenblatt1958} mechanism is the update rule
\begin{equation}
 s_i'=f(x_i),\mbox{ with } x_i = \sum_{j=1}^nw_{ij}s_j-\theta_i,
\end{equation}
which determines the probability $s_i'$ of the neuron being active. This rule involves an activation function $f(x)$, the network topology induced by the weights $w_{ij}$ and the intrinsic biases $\theta_i$. When the activation $f(x)$ is a step function, the neuron's response is bistable and reproduces the McCulloch and Pitts \cite{McCulloch1943} model. However, it is more interesting to work with sigmoid functions ---e.g. the logistic function $f(x_j)=1/(1+e^{-x_j})$ in Fig.\ \ref{fig:network}a---, because they satisfy the conditions of the ``universal approximation theorem''\ \cite{Hornik1989}. More precisely\ \cite{Note1}, any continuous function of $N$ input bits $Q(s_1,\ldots,s_N)$, can be approximated using the response of $M$ additional neurons to those input bits, as $Q\simeq \sum_{k=N+1}^{N+M}\alpha_k s_k'$. The weights $\alpha$ and $w$, and the biases $\theta$, can be optimized or \emph{trained} to minimize the approximation error, even when we ignore the function $Q$, such as in data classification and inference tasks. Even though the universal approximation theorem only requires two layers, the power of neural networks can be significantly enhanced using deep, nested architectures with multiple hidden layers. In particular, the final sum of the approximation theorem can be perfomed by one neuron, as shown in Fig.\ \ref{fig:network}b, with $w \propto \alpha$, to reconstruct the output function $Q \propto s_{final}'.$


\section{Quantum perceptron}%
Similar to Ref.\ \cite{Cao2017}, we implement a perceptron as a qubit that undergoes a SU(2) rotation\ \eqref{eq:quantum-neuron} parameterized by an external input field $\hat{x}_j$:
\begin{align}
  \label{eq:Uj}
  \hat U_j(\hat{x}_j;f) &= \exp\left\{i \arcsin[f(\hat{x}_j)^{1/2}] \hat\sigma^y_j\right\}\mbox{ with},\\
  \hat{x}_j & =\sum_{k<j} w_{jk} \hat\sigma^{z}_k - \theta_j.\nonumber
\end{align}
The perceptron qubit is characterized by quantum observables $(\hat\sigma^x,\hat\sigma^y,\hat\sigma^z)$ that rotate as
\begin{align}
  \label{eq:concatenate} \hat{\sigma}_{j}^{z\prime}= U^\dagger_j\hat\sigma_j^zU_j &=C(\hat{x}_j)\hat\sigma_{j}^{z}+S(\hat{x}_j)\hat\sigma_{j}^{x}, \\
  \hat{\sigma}_{j}^{x\prime}=U^\dagger_j\hat\sigma_j^xU_j&=-S(\hat{x}_j)\hat\sigma_{j}^{z}+C(\hat{x}_j)\hat\sigma_{j}^{x},\notag
\end{align}
and $\hat\sigma^{y\prime}_j=\hat\sigma^y_j$, with nonlinear functions $C(\hat{x}_j)=1-2f(\hat{x}_j)$, $S(\hat{x}_j)=2\sqrt{f(\hat{x}_j)[1-f(\hat{x}_j)]}$, that depend on the quantum field $\hat{x}_j$ generated by earlier neurons. This relation can be arbitrarily nested by the application of additional perceptron gates, that entangle those perceptrons with the input neurons and with earlier perceptrons, in a \emph{deep learning} scheme. In this context, notice that when $w_{lj}\neq 0$, perceptron $l > j$ will be affected by the diagonal elements $\hat\sigma^z_j$ \emph{and} the quantum fluctuations $\hat\sigma^x_j$ of the $j$-th perceptron, adding generalization power to the network.

The quantum perceptron contains the classical neural network as a limit and therefore satisfies the universal approximation theorem\ \cite{Note1}. Let us assume a three layer setup such as the one in Fig. \ref{fig:network}b, with the following conditions: (i) we have $N$ input qubits, $M$ internal perceptrons and $1$ output perceptron; (ii) all perceptrons are initially in the unexcited state and the input layer is initialized to a classical input, $\ket{s_1,s_2\ldots s_N}\ket{0_{N+1}\ldots 0_{N+M}0_{N+M+1}},$ (iii) the final perceptron's weights and threshold are tuned to explore only the linear part of the sigmoid activation function $f(x)\propto x.$ Then, the output perceptron will be excited with a probability $s_{out}=\frac{1}{2}(\braket{\hat\sigma^z_{N+M+1}}+1)$
\begin{align}
s_{out}\simeq \sum_{j=1}^M w_{N+M+1,N+j} \braket{f\Big(\sum_{k=1}^Nw_{N+j,k}s_k-\theta_{N+j}\Big)}.
\end{align}
This output probability is a linear combination of sigmoid functions of the input neurons: by virtue of the universal approximation theorem, this implies that $s_{out}$ can be used to approximate any function $Q(\hat\sigma^z_1,\ldots,\hat\sigma^z_N)$ of the input neurons\ \cite{Note1}. This is true even when we do not measure the intermediate neurons ---indeed, measuring those neurons introduces shot noise in the estimate of $\hat\sigma^z_{N+M+1}$, deteriorating the approximation.

\begin{figure}[t]
\centering
\includegraphics[width=0.6\linewidth]{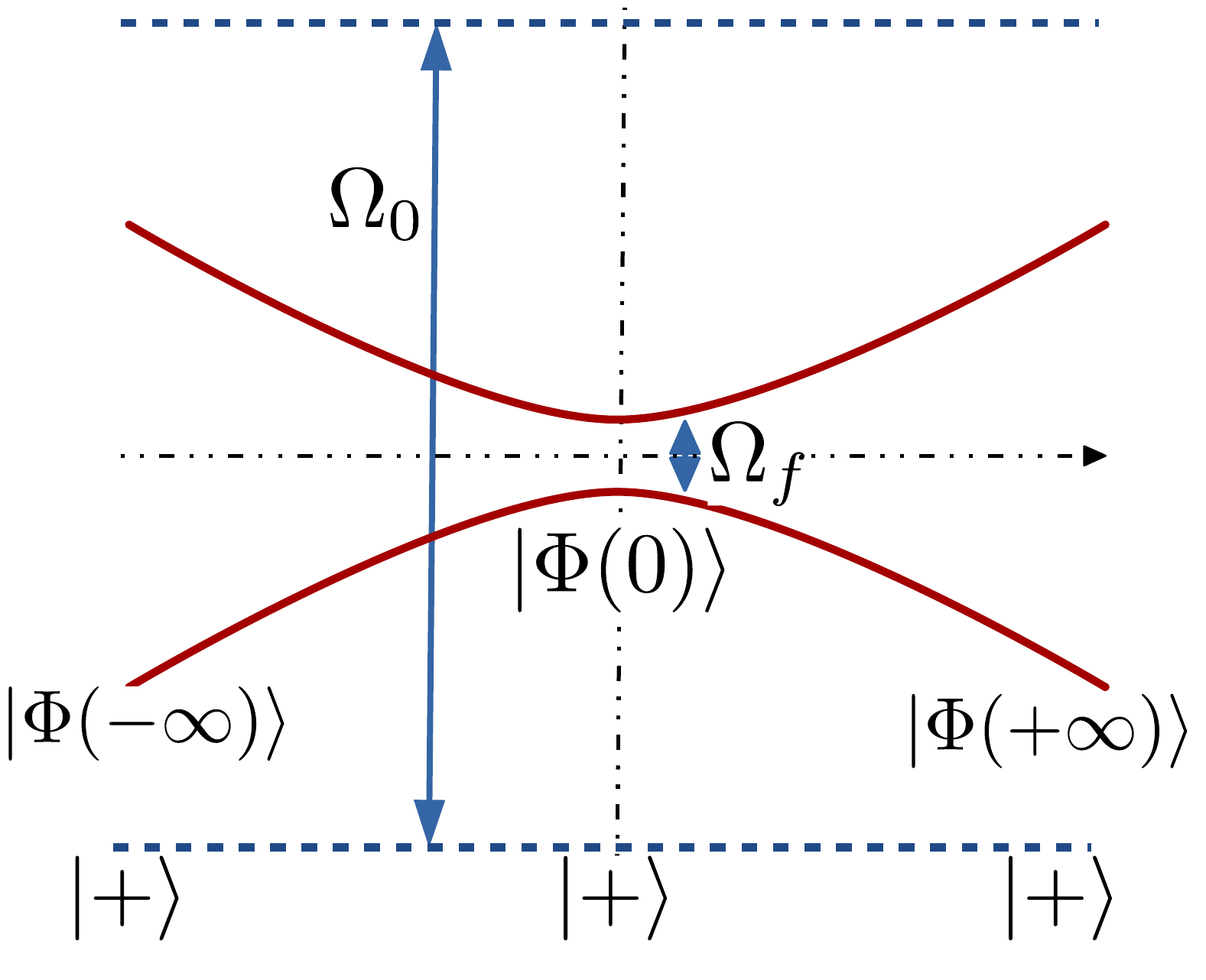}
\caption{Energy levels of the two-level system \eqref{Hqnn} as a function of the activation potential $x_j$. The perceptron gate begins with large transverse field, $\Omega_0\gg |x_j|$, such that the ground state is the approximate superposition $\ket{+}\propto \ket{0}+\ket{1}$. When the transverse field is decreased, the state converges to $\ket{\Phi(x_j/\Omega_f)}$ given by\ \eqref{eq:ideal-state}.
\label{fig:passage}}
\end{figure}

\section{Implementation}%
The second and most important result in this work is that the perceptron gate can be implemented as a single (fast) adiabatic passage in a model of interacting spins, which opens the door to specialized hardware implementation of the quantum neural network. We construct the perceptron gate evolving a qubit with the Ising Hamiltonian
\begin{eqnarray}
\label{Hqnn}
  \hat{H}(t) &=& \frac{\hbar}{2}\left[-\Omega(t)\hat{\sigma}^x_j - \hat{x}_j \hat{\sigma}_j^z\right] \nonumber\\
  &=&  \frac{\hbar}{2}\left[-\Omega(t)\hat{\sigma}^x_j+\theta_j\hat{\sigma}_j^z- \sum_{k<j} \big(\omega_{jk}\hat\sigma_k^z\hat\sigma_j^z\big)\right].
\end{eqnarray}
The qubit is controlled by an external transverse field $\Omega(t)$, has a tuneable energy gap and interacts with other neurons through $\hat{x}_j.$ The instantaneous ground state of this Hamiltonian
\begin{equation}
  \label{eq:ideal-state}
  \ket{\Phi(\hat{x}_j/\Omega)} = \sqrt{1-g(\hat{x}_j/\Omega)}\ket{0}+\sqrt{g(\hat{x}_j/\Omega)}\ket{1}
\end{equation}
has a sigmoid excitation probability [cf. Fig.\ \ref{fig:network}a, solid]
\begin{equation}
\label{gx}
g(x)=\frac{1}{2}\left(1+x/\sqrt{1+x^2}\right).
\end{equation}
This suggests implementing the gate\ \eqref{eq:quantum-neuron} in three steps: (i) set the perceptron to the superposition $\ket{+}=\mathcal{H}\ket{0}=\frac{1}{\sqrt{2}}(\ket{0}+\ket{1})$ with a Hadamard gate; (ii) instantaneously boost the magnetic field $\Omega(0)=\Omega_0\gg |\hat{x}_j|$; (iii) adiabatically ramp-down the transverse field $\Omega(t_f)=\Omega_f$ in a time $t_f$, to do the transformation $\mathcal{A}(\hat{x}_j)\ket{+}\simeq \ket{\Phi(\hat{x}_j/\Omega_f)}$.

As sketched in Fig.\ \ref{fig:passage}, the energy gap in this protocol is larger than $|\Omega(t)|$, ensuring many quasiadiabatic strategies $\Omega(t)$ to approximate $\hat U_j(\hat{x}_j;g)\simeq\mathcal{A}(\hat{x}_j)\mathcal{H}$ for $|\hat{x}_j|\leq |\hat{x}_\text{max}|\ll|\Omega_0|$. We designed two: a linear ramp $\Omega(t) = \Omega_0(1-t/t_f)+\Omega_f t/t_f$, and a FAQUAD (Fast-Quasi-Adiabatic passage) control\ \cite{Martinez-Garaot2015} that limits non-adiabatic errors\ \cite{Note1}. As figure of merit we use the average fidelity
\begin{equation}
  \bar{\mathcal{F}} = \int_{-x_\text{max}}^{x_\text{max}} \mathcal{F}[\Phi(x_j),\phi(t_f,x_j)]\mathrm{d}x_j.
\end{equation}
with $\mathcal{F}(\Phi,\phi) = |\braket{\Phi(\hat{x}_j/\Omega_f)|\phi}|^2$ and $\phi$ the final dynamical state driven by $\Omega (t)$.

\begin{figure}[t]
  \centering
  \includegraphics[width=.7\linewidth]{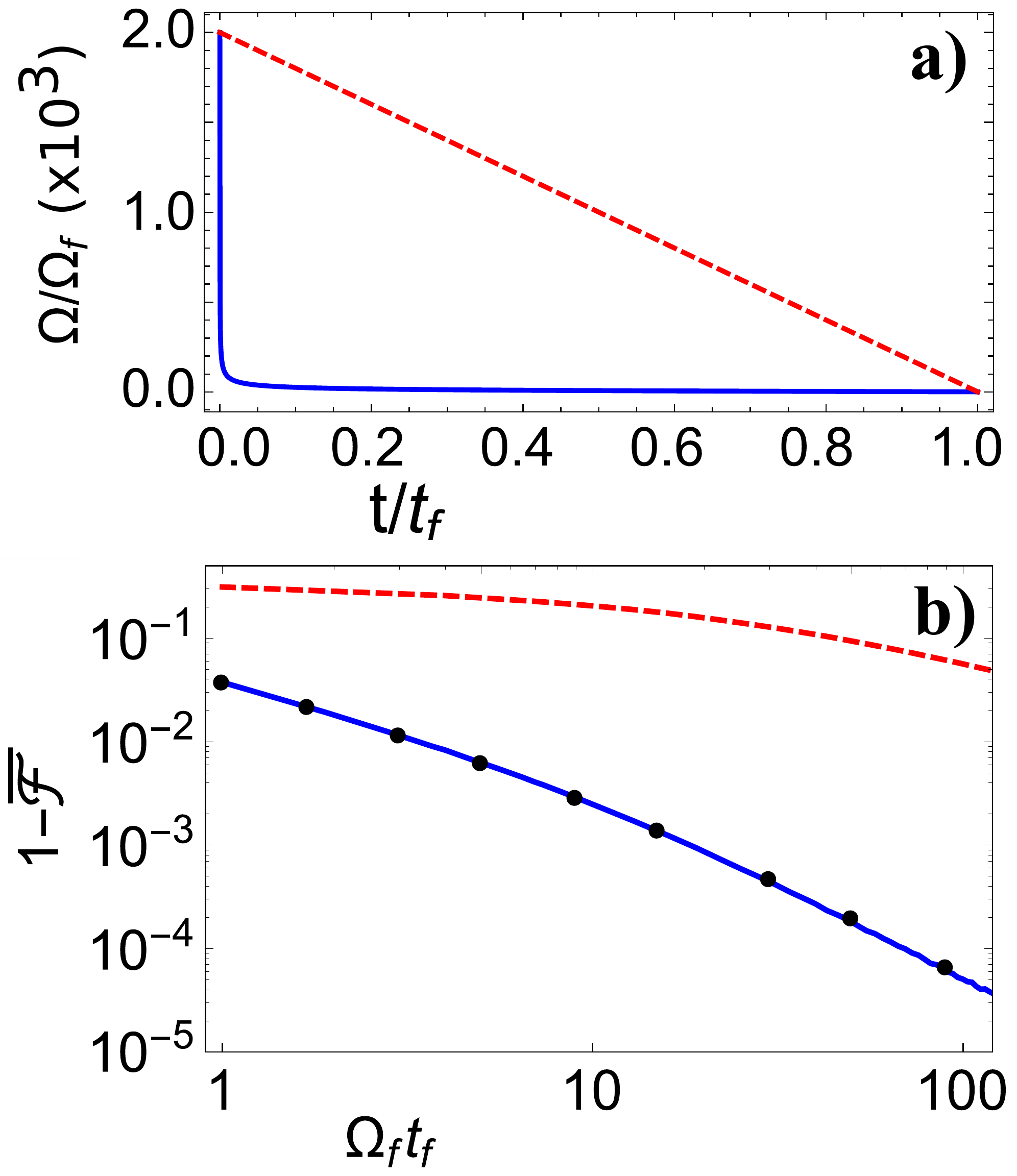}
  \caption{(a) Transverse field $\Omega(t)$ for the linear ramp (dashed) and FAQUAD (solid) protocols to implement the perceptron gate. (b) Average infidelity $1-\mathcal{\bar{F}}$ as a function of the total ramp time $t_f$, for the two ramp protocols. The FAQUAD process is fitted by $\sim c_0\exp[-c_1(\Omega_ft_f)^{c_2}]$, $c_0=26.838, c_1= 6.577$, and $c_2=0.150$ (black circles). }
\label{control}
\end{figure}

Figure\ \ref{control}a compares the linear and FAQUAD strategies to modify the transverse field. In Fig. \ref{control}b we observe that for the same time $t_f$ the FAQUAD protocol is more accurate; alternatively, given an error tolerance $\varepsilon=1-\mathcal{\bar{F}}$, the FAQUAD design is 2-3 order of magnitudes faster than the linear ramp. The quantum perceptron also shows robustness against non-adiabatic passages and high deviations, beyond experimental errors, when scheduling the control \cite{Note1}. From approximate fits and using the adiabatic passage as reference, see Fig. \ref{control}b, we estimate that the total time for a perceptron gate to have an error $\varepsilon$ scales as $t_{f,\varepsilon}=\mathcal{O}(\log(\varepsilon)^{1/0.15}\Omega_f^{-1})$. When we have multiple neurons $N$ spread over $L$ layers, the gates of a single layer can be parallelized, keeping the total time bounded, but control errors accumulate exponentially with the number of qubits. A more realistic scaling that takes this into account is $T_{f,\varepsilon}=\mathcal{O}(L\times\log(\varepsilon/N)^{1/0.15}\Omega_f^{-1})$. 

We can compare this performance with a proposal for implementing a quantum perceptron using auxiliary qubits, conditioned rotations and measurements\ \cite{Cao2017}. The gate implemented in that work is a rotation $\hat U = \exp[iq^{(k)}(x)\hat\sigma^y]$ with a nonlinear angle $q^{(k)}(x) = 2\arctan[\tan^{2^k}(x)]$ that converges to a step-wise function in the interval $x\in [-\pi/4,\pi/4]$. This gate requires about $k$ auxiliary qubits, a circuit depth $\mathcal{O}(14^k)$ and the total gate time scales polynomially $\mathcal{O}((n/\delta)^2)$ with the number of neurons per layer $n$ and the step width $\delta\simeq \Omega_f$ of the network. An important point in the work by Cao\ \textit{et al} is that it demonstrates algorithmic applications for neural networks that are perfectly discriminating ---rotation angles take values close to $\pi/2$ or $0$ and $P_j$ is either 0 or 1, as in the McCulloch and Pitts \cite{McCulloch1943} model---: those applications can also be reproduced with our Ising model perceptron by a suitable design of the final transverse field $\Omega_f$ and the biases $\theta_j$. Finally, we have to remark that our perceptron's sigmoid response is easily tuned ---the step size of $q^{(k)}$ only takes fixed values $\simeq 2^{-k}$---, and it does not have wraparound problems. These advantages are relevant for broader applications such as sensing of unconstrained input fields $x$ and are required for the perceptron to approximate arbitray operations.

\section{Parameterized quantum control}%
The quantum perceptron is an instance of a new problem in optimal control theory \cite{Palao2002, Torrontegui2013}: to design a family of unitary operations that depend on a single parameter $\hat U_x: x\in[-x_\text{max},x_\text{max}]\to \text{SU}(2)$, using a single control $\Omega(t)$ that does not have any knowledge of this parameter. The closest problem that we know of appears in NMR protocols for suppressing decoherence\ \cite{Ruschhaupt2012, Daems2013, Levy, Levy2017}: the external field $x$ is created by an environment or residual cross-talk, and the goal is to preserve the quantum state $\hat U_x\sim 1$ or do the same unitary operation for any $x$. However, the quantum perceptron is far more general and includes other multiqubit gates.

For instance, the quantum perceptron can achieve multiqubit conditional quantum gates that have the form $\hat{W}_{mqb}=\exp[i Q(\hat\sigma^z_1,\ldots,\hat\sigma^z_{j-1})\hat\sigma^y_j],$ with general continuous activation functions $Q$. The idea is to decompose the function $Q$ as a linear combination of sigmoid excitation profiles $Q(\hat\sigma^z_1,\ldots,\hat\sigma^z_{j-1})\sim \sum_n \arcsin[f(\sum_{k<j}w_{jk}^{(n)}\hat\sigma^z_k-\theta_j^{(n)})]$, reconstructing the multiqubit gate by several applications of perceptron gates with different parameters
\begin{align}
  \label{eq:combination}
  \hat W_{mqb} \simeq \prod_n \hat U_j(\sum_{k<j}w_{jk}^{(n)}\hat\sigma^z_k-\theta_j^{(n)}; f).
\end{align}
Take for instance a XOR-like gate that flips a bit when the number of excited qubits are within a given range
\begin{equation}
  s_{N+1} \to \bar{s}_{N+1}\; \mbox{if}\; M_1 < \sum_{i=1}^N s_i < M_2.
  \label{general-XOR}
\end{equation}
The ordinary XOR gate has $N=1$ input, and thresholds $M_1=0$ and $M_2=2$, but cannot be implemented using a single classical perceptron\ \cite{Minsky1969}. We can nevertheless implement the conditional logic\ \eqref{general-XOR} quantum mechanically, using two adiabatic passages with two different gaps $\theta_j^{(n)}$ and opposite signs of $\Omega_{0,f}$ for each passage, thus achieving upper and lower excitation thresholds [cf. in Fig.\ \ref{fig:other}].

\section{Quantum sensing}
\begin{figure}
  \centering
  \includegraphics[width=0.9\linewidth]{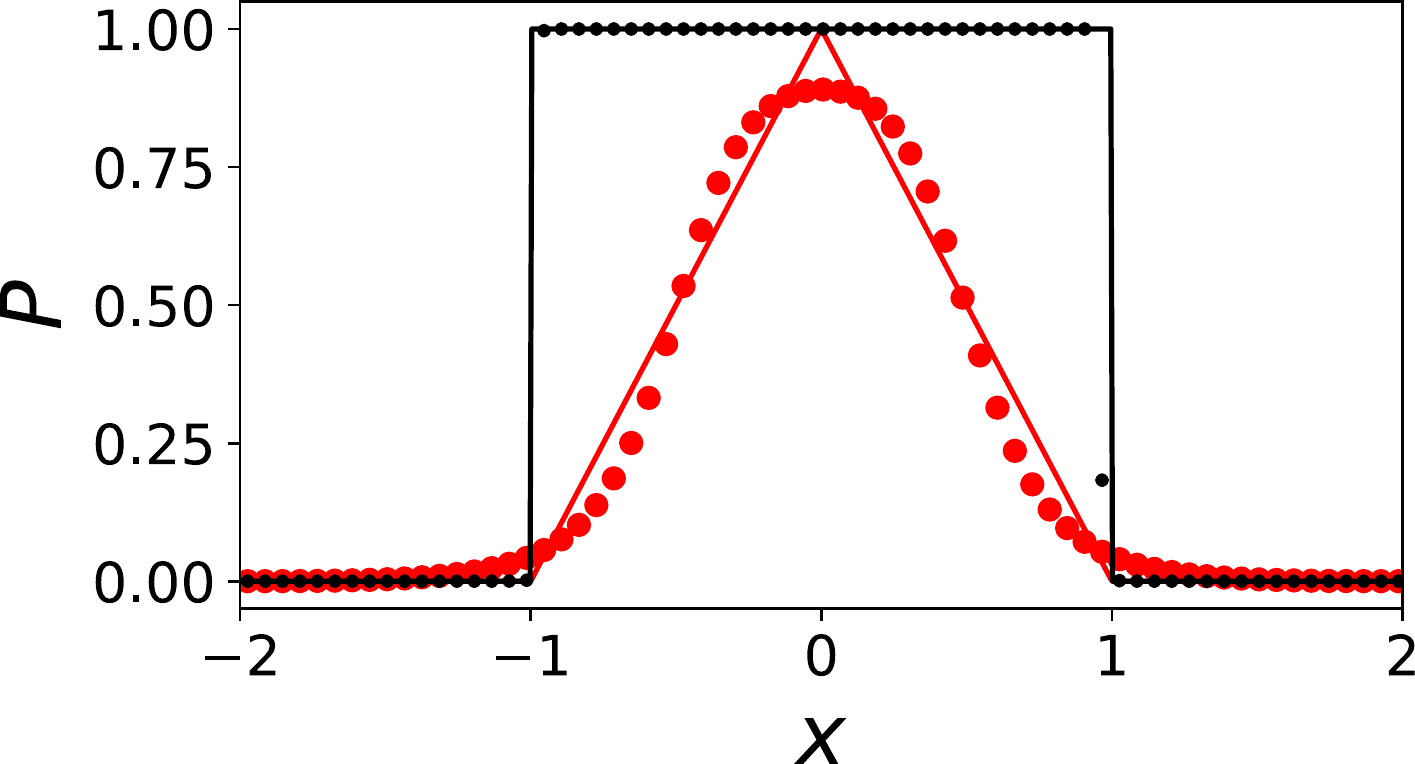}
  \caption{Perceptron responses that result from two applications of the nonlinear gate with different shifts and widths: we show ideal, non-differentiable curves (solid) and optimized fits to the gate (dots) following Eq.\ \eqref{eq:combination}.}
  \label{fig:other}
\end{figure}

Using the same ideas as for the design of multiqubit gates, we can engineer quantum sensors with resposes that go beyond interference patterns. Such sensors would overcome the problems of phase wrapping, working as threshold- or range-sensors. As examples, Fig.\ \ref{fig:other} shows two possible activations that are reconstructed with just two cycles of  the perceptron gates: the rectangular shape [cf. Fig.\ \ref{fig:other}] required for the XOR gate\ \eqref{general-XOR}, and a peaked response. Both examples were created using machine learning training algorithm in Tensorflow, recognizing that the product of unitaries in Eq.\ \eqref{eq:combination} can be written as a single exponential where the rotation angle is an instance of a neural network.

Another application of the perceptron gate would be to reconstruct global properties from the signals sensed by multiple quantum sensors. Let us assume that we have an object with a property $\chi$ ---a dipolar moment, a quadrupolar moment, a charge, etc---. This object is the source of an electromagnetic field $\phi(x,t;\chi)$ that is ultimately detected by a set of $N$ quantum sensors, whose state is changed: $ \hat{\sigma}_{n}^{z\prime} \to \hat U_n^\dagger \hat\sigma^z_n \hat U_n,$ with $\hat U_n=\exp[-i\phi(x_i,t;\chi)\hat\sigma^y_n].$ If the sensors are initially polarized, all in the same state, there will be a mapping between the values of the transformed $\hat\sigma^z_n$ to the desired property. In other words, $\tilde\chi \simeq Q(\hat\sigma^z_1,\ldots,\hat\sigma^z_n).$ This suggests adopting a scheme such as the one in Fig.\ \ref{fig:network}b, where the first layer would be the sensors and the final qubit will provide an approximation of the detected property $s_{out}\simeq \tilde{\chi}.$ Note that, by not measuring neither the sensors nor the intermediate qubits, we achieve an enhanced sensitivity with respect to a classical estimate $\tilde\chi \simeq Q(\braket{\hat\sigma^z_1},\ldots,\braket{\hat\sigma^z_n}).$

\section{Conclusion}%
Summing up, we have introduced a quantum perceptron as a two-level system that exists in a superposition of resting and active states, and which reacts nonlinearly to the field generated by other neurons. When combined with other perceptrons in a neural network configuration, this nonlinear transformations acts as a universal approximator of arbitrary computable functions, and generator of sophisticated multiqubit operations beyond the M\o{}lmer-S\o{}rensen gate \cite{Molmer1999}. In the Supplementary Material\ \cite{Numerical} we attach sophisticated numerical files to construct, train, and illustrate the approximation power and nesting of quantum perceptrons, training classically a small quantum network to detect prime numbers.

The second most important result is an implementation of the quantum perceptron gate as a quasiadiabatic passage on an Isign-type spin model. The resources in this implementation  scale favorably with the network size and the total circuit error, and the adiabatic procedure has already been demonstrated in highly connected architectures with superconducting qubits\ \cite{Lucero2012, Boixo2014, Pudenz2014}, trapped ions\ \cite{Brown2016, Graß2016} and nuclear magnetic resonance\ \cite{Steffen2003}.

The perceptron gate is a multiqubit primitive that can be integrated in quantum computing environments ---as primitives for the approximation of general discrete functions, as approximate classifiers of complex datasets, as implementation of a quantum oracle---. The model of a quantum perceptron that we have introduced has other important ramifications, such as the design of complex controlled operations or the connection to quantum sensing sketched above. In particular, the image of the multi-layer perceptron circuit as a quantum sensor opens many interesting questions. For instance, how to define and optimize the sensitivity of these sensors? Can these threshold sensors be combined with other unitary operations, quantum states, etc? If so, what are the quantum limits of threshold sensing vs. ordinary sensing of classical fields? We expect to address these and other questions in future works.

\begin{acknowledgements}
We acknowledge funding from MINECO/FEDER Project FIS2015-70856-P, CAM PRICYT project QUITEMAD+CM S2013-ICE2801, and Basque Government (Grant No. IT986-16).
\end{acknowledgements}

%
%

\bibliographystyle{apsrev4-1}
\bibliography{qnn.bbl}

\clearpage

\begin{widetext}
  \centering
  \large\textbf{Supplementary material for ``Unitary quantum perceptron as efficient universal approximator''}
\end{widetext}

\setcounter{equation}{0} \setcounter{figure}{0} \setcounter{table}{0}
\setcounter{page}{7} \makeatletter \global\long\def\theequation{S\arabic{equation}}
\global\long\def\thefigure{S\arabic{figure}}

\section{Universal approximation theorem for Classical Neural Networks}
\label{app:universalCNN}

The capacity and versatility of neural networks to classify complex data relies in the ``universal approximation theorem'', we now recall it in the form by Cybenko \cite{Cybenko1989}.

\begin{theorem}
  \label{th:classical-approximation}
  Let $I_N=[0,1]^N$ be the $N$-dimensional unit cube and $C(I_N)$ the space of continuous functions on $I_N$. Let the function $\eta$ be continuous and sigmoidal ---i.e. $\eta(\infty)\to 1$, $\eta(-\infty)\to 0$---. Then, finite sums of the form
  \begin{equation}
  \label{Cyb}
    q(\vc{s}) = \sum_j^{M_\varepsilon} \alpha_j \eta\left(\sum_{k=1}^N w_{jk}s_k - \theta_j\right)
  \end{equation}
  are dense in $C(I_N)$. In other words, given any $Q\in C(I_N)$ and $\varepsilon>0$, there exists a sum $q(\vc{s})$ with $M_\varepsilon$ terms, for which $|Q(\vc{s})-q(\vc{s})|\leq \varepsilon$ for all $\vc{s}\in I_N$.
\end{theorem}

\noindent
Following this theorem, we can design a three-layer neural network with $N$ input, $M_\varepsilon$ intermediate or ``hidden'' neurons and a single ``output'' neuron, to approximate any function $Q(\vc{s})\in C(I_N)$. The $N$ input neurons will be assigned the argument of the function we wish to compute $s_{\text{in},i} = s_i$. We will use the graded response update to determine the values of the $M_\varepsilon$ hidden neurons $s_{\text{hid},j}=f(\sum_{k=1}^N w_{jk}s_{\text{in},k}-\theta_j),\; j=1,\ldots,M_\varepsilon$. Finally, we will collect all the values in a final neuron, working close to the linear regime
\begin{align}
  s_\text{final} &=
                   \eta\left(\sum_{j=1}^{M_\varepsilon} \alpha_j s_{\text{hid},j}\right)
  \simeq
  \sum_{j=1}^{M_\varepsilon} \alpha_{j}s_{\text{hid},j}
\end{align}
and approximate $Q(\mathbf{s})\simeq s_\text{final}.$
Determining the values of $\alpha_j$, $w_{jk}$ and $\theta_j$ for an specific function amounts to training the network.

\begin{figure}[h]
  \centering
  \includegraphics[width=0.7\linewidth]{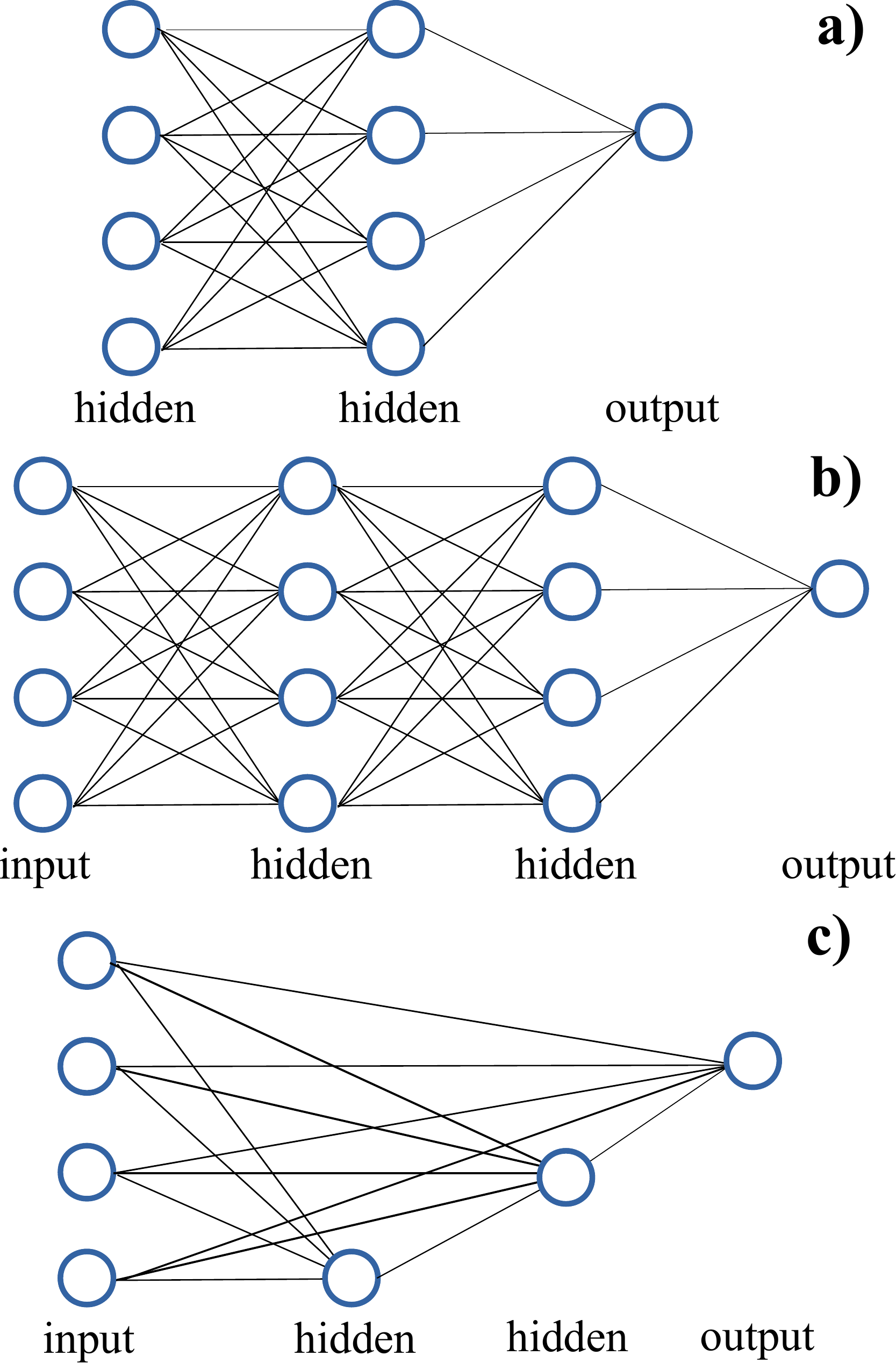}
  \caption{(a) The neural network of the universal approximation theorem consists of three layers. The input layer contains the arguments of the function $Q$ that we wish to approximate. The output layers recreates the approximation using sigmoid response functions. The final network collects all the output neuron values and adds them up according to Eq.\ \eqref{eq:real-approx}. However, the quantum perceptron can be used in other topologies with deeper nesting (b), or even with transverse dependencies between layers (c), allowing for quantum versions of \emph{deep learning.}}
  \label{fig:deep}
\end{figure}

\section{Universal approximation theorem for Quantum Neural Networks} \label{universalQNN}

\begin{theorem}
  \label{th:approximation}
  Any bounded continuous function $Q(\hat\sigma_1,\ldots,\hat\sigma_N) \in [-1,1]$ of the quantum observables $\{\hat\sigma_i\}_{i=1}^N$ can be reconstructed up to an error $\varepsilon$ onto the state of a qubit using $N$ input qubits and $M_\varepsilon+1$ applications of the quantum perceptron gate.
\end{theorem}

The proof relies on Theorem \ref{th:classical-approximation} and on the transformation implemented by the quantum perceptron gate in the Heisenberg picture
\begin{align}
  \hat{\sigma}^{z\prime}= U^\dagger\hat\sigma^z U &=C(\hat{x})\hat\sigma^{z}+S(\hat{x})\hat\sigma^{x}.\notag
\end{align}
with nonlinear functions $C(\hat{x}_j)=1-2f(\hat{x}_j)$, $S(\hat{x}_j)=2\sqrt{f(\hat{x}_j)[1-f(\hat{x}_j)]}$ and perceptron sigmoid response
\begin{equation}
  f(x) = \frac{1}{2}\left(1+x/\sqrt{1+x^2}\right).
\end{equation}

We will repeat the same structure of the classical neural network in Fig.\ \ref{fig:deep}, where we have $N$ input neurons, $M_\varepsilon$ hidden qubits that build the approximation and a final qubit that collects the information. The $M_\varepsilon+1$ qubits will be set in the initial state $\ket{0}$. The $M_\varepsilon$ qubits on the first layer will transform into
\begin{align}
  \hat{\sigma}_{\text{hid},j}^{z\prime} &=C(\hat{x}_j)\hat\sigma^{z}_{\text{hid},j}+S(\hat{x}_j)\hat\sigma^{x}_{\text{hid},j}\notag,\;\mbox{with}\\
  \hat{x}_j & = \sum_k w_{jk}\hat\sigma^z_{\text{in},k}-\theta_j.
\end{align}
We tune the final neuron to work in the linear regime of the sigmoid function. Defining $\hat{x}_\text{out}= \sum_j \alpha_j \hat\sigma^{z\prime}_{\text{hid},j}-\theta_\text{out}$, we will require $C(\hat{x}_\text{out})=1-2f(\hat{x}_\text{out})\simeq 1- 2\hat{x}_\text{out}.$ Then
\begin{align}
  \hat{\sigma}_{\text{out}}^{z\prime} &
  \simeq\left( 1-2 \sum_j\alpha_j \hat\sigma^{z\prime}_{\text{hid},j}+2\theta_\text{out} \right)\hat\sigma^{z}_{\text{out}}+S\hat\sigma^{x}_{\text{out}}\\
 &=\left\{1+2\theta_\text{out}-2\sum_j\alpha_j [1 - 2f(\hat{x}_j)]\hat\sigma^z_{\text{hid},j}\right\}\hat\sigma^z_\text{out}\notag \\
 & + \hat O_{corr}. \notag
\end{align}
In this last line we introduced a new operator $\hat O_{corr}$ which contains corrections that are proportional to $\hat\sigma^x_{\text{hid},j}$ and $\hat\sigma^x_\text{out}.$ Since the initial state of the protocol is an eigenstate of all the $\hat\sigma^z_{\text{hid},j}$ operators, when we compute the total excitation probability of the output neuron $s_\text{out}= \frac{1}{2}(1+\braket{\hat\sigma^{z\prime}_\text{out}})$, we find
\begin{align}
\label{eq:real-approx}
  s_{out}&=\braket{\left\{1+\theta_\text{out}-\sum_j\alpha_j [1 - 2f(\hat{x}_j)]\hat\sigma^z_{\text{hid},j}\right\}\hat\sigma^z_\text{out}}
           \notag
  \\&=\braket{\left\{1+\theta_\text{out}+\sum_j\alpha_j [1 - 2f(\hat{x}_j)]\right\}(-1)}
  \notag
  \\
         &=\sum_j2\alpha_j \braket{f\Big(\sum_kw_{jk}\hat\sigma^z_{\text{in},k}-\theta_j\Big)}
  \\
  &\simeq  \braket{Q(\hat\sigma^z_{\text{in},1},\ldots\hat\sigma^z_{\text{in},N})}.\notag
\end{align}
Here we have imposed
\begin{equation}
  1 + \theta_\text{out} +\sum_j \alpha_j = 0,
\end{equation}
and used the universal approximation theorem to find the $\alpha_j$ to approximate our generic continuous function $Q$.

It is important to remark that the previous demonstration focuses on finding a classical limit for the approximation, neglecting the quantum fluctuations that are present in $\hat{O}_\text{corr}.$ However, this is not needed for a general operation of the quantum perceptron, which may benefit from those fluctuations to implement more general approximants than those of a classical feed-forward neural network. In particular, section {\itshape Deep Learning} discusses a procedure to optimize an approximation with a quantum neural network of arbitrary depth which does not require any of the previous constraints.

\section{Fast quasiadiabatic dynamic}

Given the same boundary values $\Omega_0$ and $\Omega_f$ as for a linear ramp, we can engineer a rather fast  control of $\Omega(t)$ that still achieves the target state (8) for all $x_j$. The need to produce single controls independently on one Hamiltonian parameter automatically discards many of the existing methods that speed up adiabatic passages\ \cite{Torrontegui2013, Palao2002}. However, there is one strategy of fast quasiadiabatic dynamics (FAQUAD)\ \cite{Martinez-Garaot2015}, which only works with the adiabatic parameter $\mu(t)$
\begin{equation}
\label{mu}
\mu(t)=\hbar\left|\frac{\langle\phi_0(t)\ket{\partial_t\phi_1(t)}}{E_1(t)-E_0(t)}\right|
\end{equation}
expressed in terms of the rate of change of the first excited state $\ket{\phi_1(t)}$ of $\hat{H}(t)$ and the energy separation between the ground and excited states, $E_1-E_0$ of a quasiadiabatic Hamiltonian. We will generalize this strategy, imposing conditions on $\mu(t)$ that are satisfied for all input fields and states of the neurons $x_j$, thereby designing the optimal controls for implementing this gate.

\begin{figure}[t!]
   \includegraphics[width=0.65\linewidth]{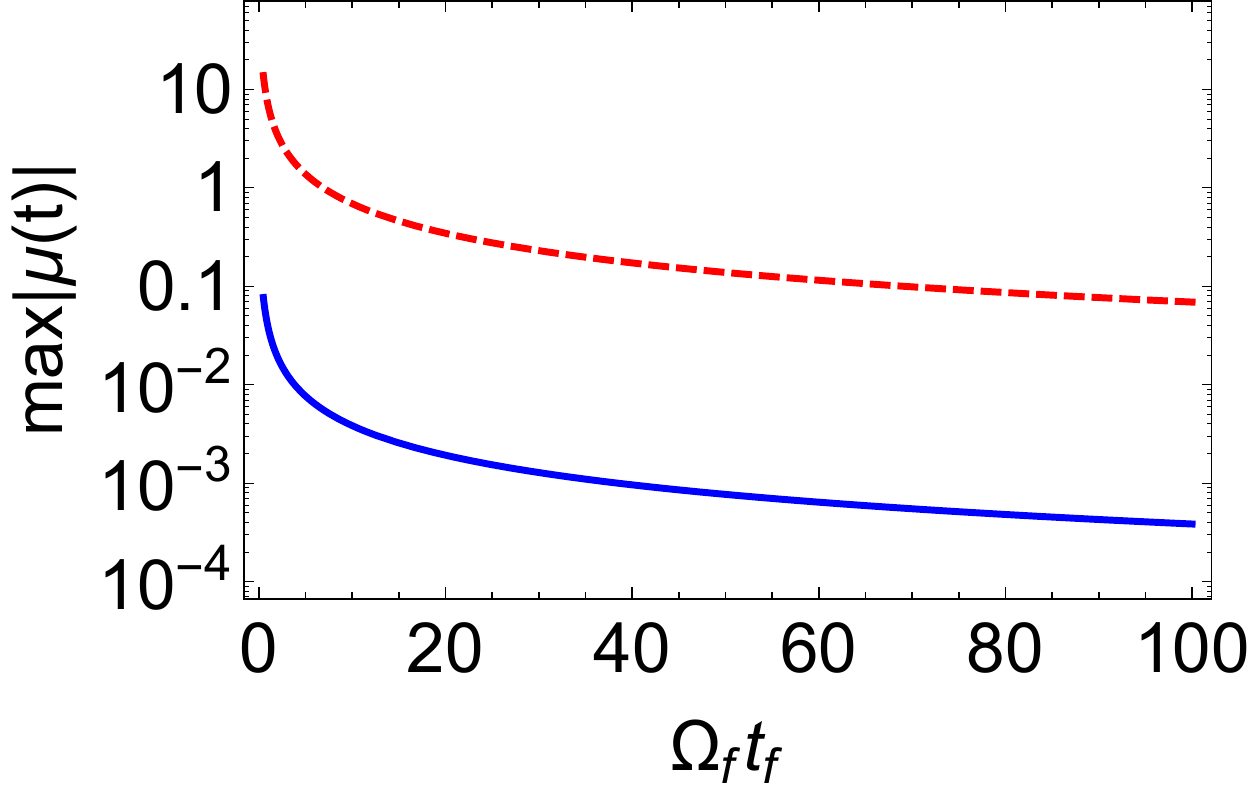}
  \caption[fig:network]{Maximum value of the adiabatic parameter for the linear and FAQUAD ramps (dashed and solid, respectively). A smaller value of $\mu(t)$ implies a lower probability of errors in the adiabatic preparation of the quantum perceptron.}
  \label{fig:mu}
\end{figure}

Our strategy will be to ensure that the adiabatic parameter remains constant $\mu(t)=c$ to delocalize the transition probability along the whole process. If the relation between field and time is invertible $t=t(\Omega)$, applying the chain rule to Eq. (\ref{mu}) gives
\begin{equation}
\label{mu_t}
\frac{d\Omega}{dt}=\pm\frac{c}{\hbar}\bigg|\frac{E_0(\Omega)-E_1(\Omega)}{\langle\phi_0(\Omega)\ket{\partial_{\Omega}\phi_1(\Omega)}}\bigg|,
\end{equation}
where the sign determines whether $\Omega(t)$ monotonously increases or decreases from $\Omega_0$ to $\Omega_f$. We rescale time according to the total duration $s=t/t_f$ and define $\tilde\Omega(s):=\Omega(s\,t_f)$ so that $d\Omega(t)/dt=t_f^{-1}d\tilde\Omega/ds.$ This way,
\begin{align}
\label{om_fac}
\frac{d\tilde\Omega}{ds}&=\pm\frac{\tilde c}{\hbar}\bigg|\frac{E_0-E_1}{\langle\phi_0\ket{\partial_{\tilde\Omega}\phi_1}}\bigg|_{\tilde\Omega},\mbox{ where }\\
\tilde c&=ct_f=\pm\hbar\int_{\tilde\Omega(0)}^{\tilde\Omega(1)}\frac{d\tilde\Omega}{\bigg|\frac{E_0-E_1}{\langle\phi_0\ket{\partial_{\tilde\Omega}\phi_1}}\bigg|_{\tilde\Omega}}.
\end{align}
To deduce $\tilde\Omega(s)$ for the FAQUAD protocol we solve Eq.\ \eqref{om_fac}, choosing $\tilde{c}$ to satisfy $\tilde\Omega(0)=\Omega_0$ and $\tilde\Omega(1)=\Omega_f$. A different election of $t_f$ corresponds to a scaling of $c=\tilde{c}t_f$ and $\Omega(t=st_f)=\tilde{\Omega}(s)$. For the particular Hamiltonian (6) the instantaneous eigenstates and energies are given by,
\begin{align}
\label{eigen}
  \ket{\phi_i}&=\cos(\theta/2)\ket{1}+(-1)^i\sin(\theta/2)\ket{0}, \\
  E_i&=-(-1)^i\sqrt{\Omega^2+x_j^2}/2, \quad i\in\{0,1\},
\end{align}
where $\theta=\arccos[-x_j/\sqrt{\Omega^2+x_j^2}]$. Replacing Eq. (\ref{eigen}) into Eqs. (\ref{om_fac}), the FAQUAD control $\Omega(t)$ is deduced. However, this transverse field
is different for different $x_j$ values. The constant adiabatic parameter for the FAQUAD protocol is
\begin{equation}
\mu=\left\vert\frac{1/\sqrt{1+x_j^2/\Omega_0^2}-1/\sqrt{1+x_j^2/\Omega_f^2}}{2x_jt_f}\right\vert.
\end{equation}
For the gate to succeed, we need a single control that does not depend on the neuron input potential $x_j$. We notice that the largest value of $|\mu|$ happens at $|x_j/\Omega_f|\approx1.272$ providing us with an optimal definition of $\mu(t)$ that works for all input neuron configurations.
In Fig. \ref{fig:mu} the corresponding adiabatic parameter $\mu$ is plotted for the linear and FAQUAD ramps as a function of $t_f$. Whereas for the FAQUAD protocol $\mu$ is constant along the whole interval, $\mu$ changes in time in the linear ramp of $\Omega(t)$ taking its maximum value at the end of the process $t=t_f$ that corresponds to the minimum energy gap $E_1-E_0$. As $t_f$ increases both protocols become more adiabatic, however, for a fixed $t_f$ value the FAQUAD strategy is more adiabatic allowing a sigmoidal excitation response in processes 2-3 orders of magnitude faster than with a simple linear ramp.

\begin{figure}[t!]
   \includegraphics[width=0.65\linewidth]{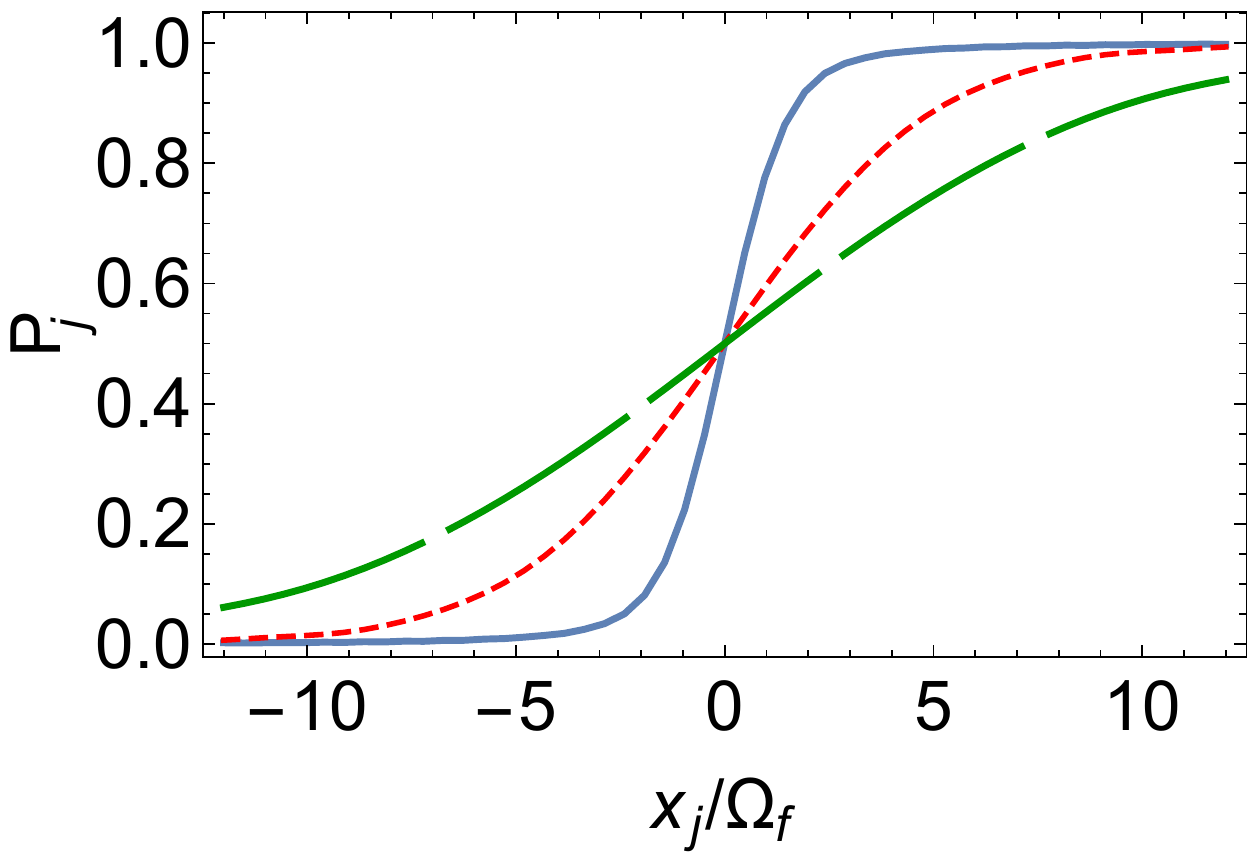}
  \caption[fig:network]{Quantum perceptron response obtained for a FAQUAD passage for different $t_f$ values. Adiabatic passage $\Omega_ft_f=10$ (solid), non-adiabatic passages  $\Omega_ft_f=1$ (short-dashed) and $\Omega_ft_f=0.25$ (long-dashed).}
  \label{fig:na}
\end{figure}

\section{Perceptron Robustness}

The universal approximation theorem relies on any univariate and sigmoidal function ---i.e. $\eta(\infty)\to 1$, $\eta(-\infty)\to 0$---, consequently the operability of the quantum perceptron is not restricted to a perfect adiabatic passage with the specific sigmoid response
given by Eq. (8). We have analyzed the robustness of the quantum perceptron against non-adiabatic passages and errors in scheduling the control.

Firstly, assuming a FAQUAD strategy the response of the perceptron for different $t_f$ values is analyzed, see Fig. \ref{fig:na}. For a perfect adiabatic protocol, blue solid line, the activation function corresponds to the algebraic sigmoid (8). For faster non-adiabatic
times $\Omega_ft_f=1$ and $\Omega_ft_f=0.25$ the response becomes flatter but it still has a sigmoidal profile that experimentally can be calibrated. 
The fidelity with respect to a perfect adiabatic passage deteriorates, see Fig. 3b, but it does not compromise the quantum perceptron operability. Sharper responses can be recovered by increasing the neuron weights $\omega_{jk}$. 

\begin{figure}[t!]
   \includegraphics[width=0.495\linewidth]{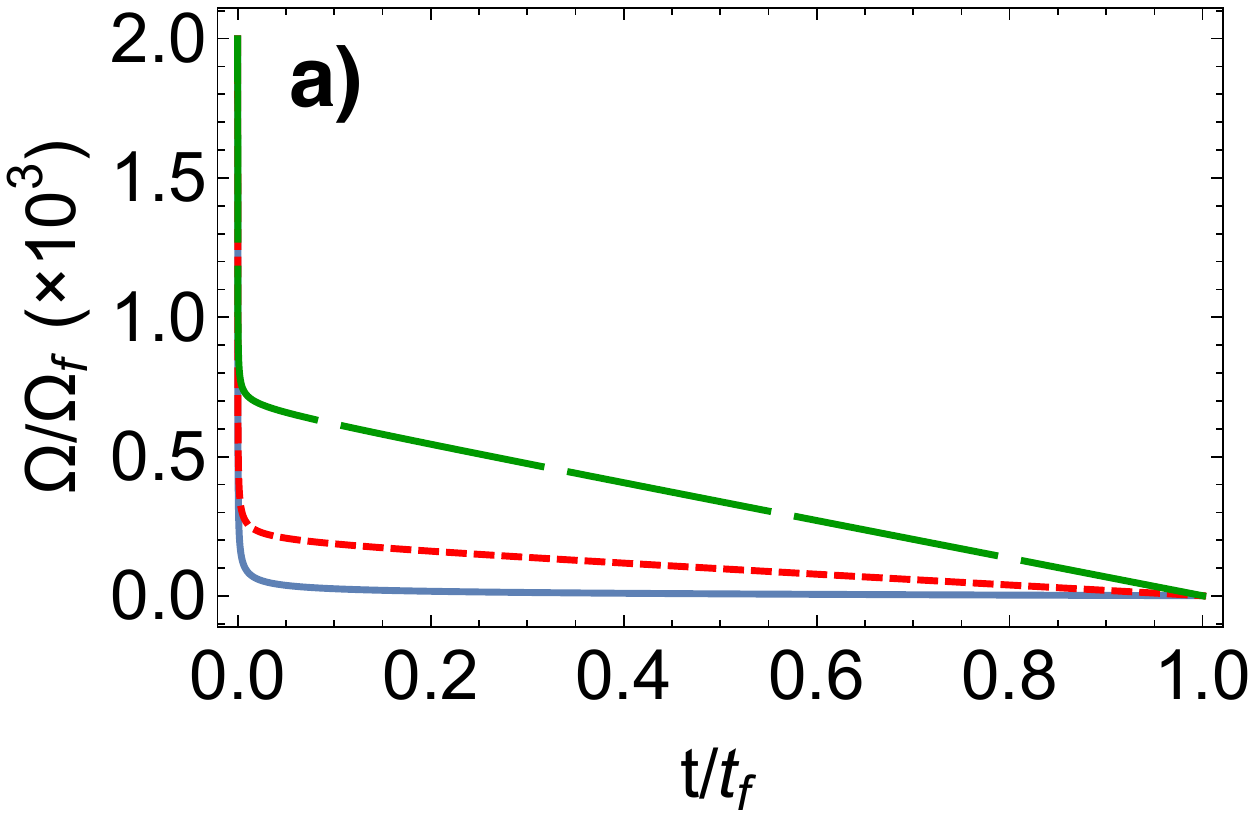}
    \includegraphics[width=0.485\linewidth]{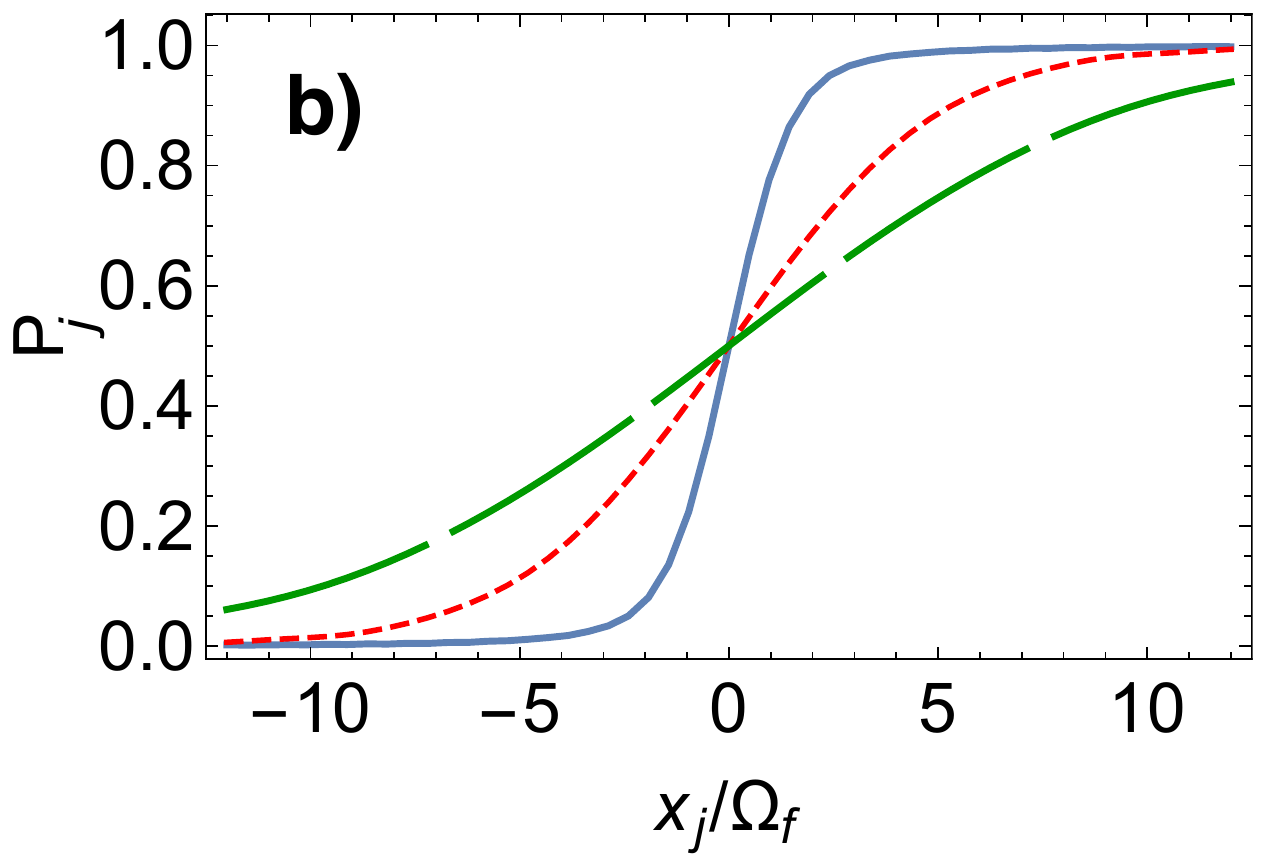}
  \caption[fig:network]{a) Different modifications of the FAQUAD control corresponding to $\epsilon=0$ (solid), $\epsilon=0.1$ (short-dashed), and $\epsilon=0.5$ (long-dashed). (b) Corresponding perceptron responses.}
  \label{fig:errors}
\end{figure}

Secondly, to analyze the effect of
errors in the implementation of the control a linear ramp $\epsilon[\Omega_0+(\Omega_f-\Omega_0)t/t_f]$ is superimposed to the FAQUAD protocol being $\epsilon$ a degradation constant. For a fixed $\Omega_f t_f=10$, Fig. \ref{fig:errors} shows the different protocols
attending to different values of $\epsilon$ and the corresponding perceptron responses. A perfect implementation of the FAQUAD control $\epsilon=0$ corresponds to the algebraic sigmoid Eq. (8) of an adiabatic passage (blue-solid line). 
Deviations in the implementation of the control, much bigger than just simple experimental errors, modify the response, however, it keeps a sigmoidal profile. As the FAQUAD passage deteriorates, $\epsilon=0.1$ and $\epsilon=0.5$ the process becomes more non-adiabatic
and the sigmoid response becomes flatter. As discussed before, it does not affect to the operability of the quantum perceptron becoming robust against both non-adiabatic passages and errors scheduling the control.

\section{Deep Learning}
\label{sec:deep}

\subsection{Deep neural network arquitecture}

The quantum perceptron can be integrated in a variety of architectures. The universal approximation theorem demands a three-layer circuit. However, state-of-the art classical neural networks work with multiple layer schemes in what is know as \emph{deep learning} [cf. Fig.\ \ref{fig:deep}b]. These architectures can also be programmed with the quantum perceptron in an efficient way.

From the main text we conclude that it is possible to implement the neuron activation gate Eq. (1) on a time $t_f=\mathcal{O}(\Omega_f^{-1})$ where $\Omega_f$ determines the interval over which the algebraic sigmoid function $g(x)$ switches from 0 to +1. If a circuit has $L$ layers, the perceptrons on the same layer can be activated simultaneously, using a generalization of the Hamiltonian (6). Labeling the neurons by layer $l$ and index within the layer and
assuming that only two consecutive layers interact simultaneously being these switched on and off forwardly, 
the Hamiltonian of the network 
\begin{align}
\hat H(t)=\frac{\hbar}{2}\sum_{l,j}\bigg[-\Omega(t)\hat{\sigma}_{j,l}^x+\theta_{j,l}\hat{\sigma}_{j,l}^z \nonumber\\
-\sum_k\omega_{jl,kl-1}\hat{\sigma}_{j,l-1}^z\hat{\sigma}_{k,l-1}^z \bigg],
\end{align}
is equivalent and uses the same control $\Omega(t)$ as for a single neuron, taking exactly the same time $t_f$. Therefore, the implementation of the total
network will have a cost $\mathcal{O}(L/\Omega_f)$ which grows polynomially with the number of layers and not with the total number of neurons.

\subsection{Classical training of the network}

We have tested the performance of the quantum perceptron network to approximate and classify various datasets. The accompanying Jupyter and Python notebooks \cite{Numerical} show a particular example where we train networks with two, three and four perceptrons to detect the prime numbers with 2 to 8 bits. Of course, these simulations are classical, but they illustrate the possibility of applying gradient methods to train the network.

The problem that we solve takes a training set of $S$ pairs containing the input and output values $\{(X_i,Y_i)\}_{i=1}^{S}$. The inputs are binary numbers $X_i=(x_{i1},x_{i2},\ldots,x_{iN})\in \mathbb{Z}_2^N$, while the outputs will be constrained into a finite interval $Y_i=Q(X_i)\in[0,1]$. In our examples $Q=1$ iff. the input number is prime. We do not know the function $Q$ but we will use $M$ perceptrons to approximate it.

The quantum approximation procedure uses $N+M$ qubits, applies $M$ perceptron gates  $\hat{U}_\text{tot} = \prod_{j=1}^M \hat U_j$ and collects the approximation into the last or output qubit $\hat\sigma^z_{N+M}.$ All gates are collectively characterized by a matrix of interactions $J_{j,k}$, the vector of thresholds $b_j$
\begin{equation}
\hat U_j = \exp\left[-i \hat\sigma^y_{N+j} \chi\left(\sum_{k < N+j} J_{j,k}\hat\sigma^z_k + b_{j}\right)\right],
\end{equation}
and the sigmoid-like excitation angle that is produced by our Landau-Zener scheme $\chi(x) = \arcsin[f(x)^{1/2}].$ Our goal will be that, given an input state $\ket{\Psi(X_i)} = \ket{x_{i1},x_{i2}\ldots x_{iN},0_{N+1},\ldots, 0_{N+M}}$, in which all perceptrons are deactivated, the output state $\hat{U}_{tot}\ket{\Psi(X_i)}$ produces a distribution 
\begin{align}
  p(X_i)
  &= \frac{1}{2}\left(  \braket{\Psi(X_i)|\hat U^\dagger \hat\sigma^z_{out}\hat U|\Psi(X_i)} + 1\right)\\
  &\simeq Y_i=Q(X_i).\notag
\end{align}
As figure of merit of the approximation we use the cross entropy $H(Y_i,p(X_i))$ between distribution $Y_i$ and the perceptron excitation probability $p(X_i)$
\begin{align}
\mathcal{C}(\mathbf{J},\mathbf{b})=&\frac{1}{S}\sum_{i=1}^SH(Y_i,p(X_i)) \nonumber\\
=&\frac{1}{S}\sum_{i=1}^S\left[Y_i\log p(X_i)+(1-Y_i)\log(1-p(X_i))\right]. 
\end{align}
The \emph{training} or the network consists in finding the matrix $\mathbf{J}$ and vector $\mathbf{b}$ that minimize the cost function $\mathcal{C}(\mathbf{J},\mathbf{b})$.

There are three important remarks to be done here. The first one is that we can reconstruct simultaneously all values of $p(X_i)$ using the unitary operation $\hat{U}_{tot}$ only once. The idea is to build the input state
\begin{equation}
  \ket{\xi} = \frac{1}{\sqrt{S}}\sum_{i=1}^S\ket{x_{i1},\ldots,x_{iN},0_{N+1},\ldots,0_{N+M}},
\end{equation}
compute $\hat U_{tot}\ket{\xi}$ and extract $p(X_i)$ from the resulting wavefunction.

The second remark is that we can use a gradient descent method to optimize the cost function $\mathcal{C}$. The simplest method, used in our example notebooks, uses the structure of the total unitary $\hat{U}_{tot}$ to compute $\partial\hat{U}_{tot}/\partial J_{n,k}$ and $\partial\hat{U}_{tot}/\partial b_n$ and then derive $\partial\mathcal{C}/\partial\mathbf{J},$ $\partial\mathcal{C}/\partial\mathbf{b}.$ The procedure used is rather straightforward, but more efficient generalizations based on backpropagation and stochastic optimization are definitely possible.

Finally, it is important to remark that we can tune the topology of the network by selecting which numbers in the connectivity matrix $J_{n,k}$ are nonzero. In our implementation the topology is fixed with a mask $\beta_{n,k}\in\{0,1\}$ such that the total connectivity matrix becomes $J_{n,k}^\text{real}=\beta_{n,k} J_{n,k}$, and we optimize for the nonzero elements $J_{n,k}^\text{real}$.

\subsection{Toy model}

As mentioned above, we have tested this approach  \cite{Numerical} using a rather complex function $Q(X_i)$, defined by a truth table that outputs $1$ iff. the number $X_i$ is prime. We have worked with problems of 3, 4, 5, 6 and up to 8 bits, constructing the function $Q(X_i)$ for each of these problems and deriving both a quantum neural network approximation and, when possible, an approximation based on classical neural networks.

It is important to remark that, in the language of machine learning, this is an overfitted example where the training and test sets coincide. It does not measure the predictive power of the quantum network, but it is perfect to demonstrate that the quantum perceptrons can approximate complex functions.

We have used two different topologies for the approximation: a three-layer network [cf. Fig. \ref{fig:deep}b] and a deep network [cf. Fig. \ref{fig:deep}c]. Out of these, the first one is also implemented with classical neural networks using Tensorflow for comparison. The first outcome of these tests is that we get good approximations and training converges to accuracies that are comparable to classical neural networks with similar topologies. The training efficiency is not that good, but this can be attributed to using global optimization methods and can be improved in the near future.

The second message is that deep neural networks with quantum perceptrons, such as in the topology of Fig. \ref{fig:deep}b, scale well and can be trained more easily to detect more numbers. This is a very promising result, because the total unitary $\hat{U}_{tot}$ used to detect prime numbers contains less parameters than other quantum algorithms that have been suggested for a similar task\ \cite{latorre14}.

\end{document}